# N-dimensional Auto-Bäcklund Transformation and Exact Solutions to n-dimensional Burgers System

Mingliang Wang [1,2], Jinliang Zhang [1*] & Xiangzheng Li [1]

1. School of Mathematics & Statistcs, Henan University of Science & Technology, Luoyang, 471023, PR China

2. School of Mathematics & Statistcs, Lanzhou University, Lanzhou, 730000, PR China

* Corresponding author. E-mail: zhangjin6602@163.com

**Abstract:** N-dimensional Bäcklund transformation (BT), Cole-Hopf transformation and Auto-Bäcklund transformation (Auto-BT) of n-dimensional Burgers system are derived by using simplified homogeneous balance (SHB). By the Auto-BT, another solution of the n-dimensional Burgers system can be obtained provided that a particular solution of the Burgers system is given. Since the particular solution of n-dimensional Burgers system can be given easily by the Cole-Hopf transformation, then using the Auto-BT repeatedly, more solutions of n-dimensional Burgers system can be obtained successively.

**Keywords:** n-dimensional Burgers system; n-dimensional BT; n-dimensional Cole-Hopf transformation; n-dimensional Auto-BT; SHB; exact solution

**AMS(2000) Subject Classification:** 35Q20; 35Q53

## 1. Introduction

The classical Burgers equation is the simplest nonlinear model in fluid dynamics[1] and has been widely used in the surface perturbations, acoustical waves, electromagnetic waves, density waves, population growth, magnetohydrodynamic waves[2-4], etc. The n-dimensional Burgers system in the form[5-7]

$$\frac{\partial u_i}{\partial t} + \sum_{j=1}^{n} u_j \frac{\partial u_i}{\partial x_j} - \mu \Delta u_i = 0 , i = 1,2,\cdots,n , \Delta \equiv \sum_{j=1}^{n} \frac{\partial^2}{\partial x_j^2}, \qquad (1)$$

with an irrotational condition:

$$\frac{\partial u_i}{\partial x_j} = \frac{\partial u_j}{\partial x_i}, \ i, \ j = 1,2,\cdots,n, i \neq j ,$$

is an important generalization of the Burgers equation and was investigated recently by Yang Chen etc. In paper [5], where the authors have shown that the n-dimensional Burgers system (1) can be transformed into n-dimensional linear heat equation by a n-dimensional Cole-Hopf transformation.

In the present paper we will study further n-dimensional Burgers system (1) by using the idea of SHB[8], which means that an undetermined function $f = f(\varphi)$ and its derivatives $f_x = f'(\varphi)\varphi_x, \ldots$, that appearing in the original homogeneous balance(HB)[9-12], are replaced by a logarithmic function $A\ln(\varphi)$ and its derivatives $A(\ln\varphi)_x = A\dfrac{\varphi_x}{\varphi}$ ,…, respectively, where constant $A$ and the function $\varphi = \varphi(x,t)$ are to be determined.

The aim of this paper is to derive the n-dimensional BT for n-dimensional Burgers system (1), and from the BT to reason out the n-dimensional Cole-Hopf transformation and the n-dimensional Auto-BT for the n-dimensional Burgers system (1). By the Auto-BT, another exact solution of the Burgers system can be obtained provided a particular solution of it is given. Hence, once the particular solution is given, using the Auto-BT repeatedly, more exact solutions of the Burgers system can be obtained successively.

## 2. Derivation of n-dimensional Auto-BT

Considering the homogeneous balance between $u_i \dfrac{\partial u_j}{\partial x_i}$ and $\dfrac{\partial^2}{\partial x_i^2} u_j$ in system (1) ($m+n+1 = n+2 \to m = 1$), according to SHB, we can suppose that the solution of the n-dimensional Burgers system (1) is of the form

$$u_i = A(\ln\varphi)_{x_i} + u_i^{(0)} = A\dfrac{\varphi_{x_i}}{\varphi} + u_i^{(0)}, i = 1,2,\cdots,n \tag{2}$$

where $u_i^{(0)} = u_i^{(0)}(x_1, x_2, \cdots, x_n, t), i = 1,2,\cdots,n$, be a given particular solution of n-dimensional Burgers system (1), i.e.

$$\dfrac{\partial u_i^{(0)}}{\partial t} + \sum_{j=1}^{n} u_j^{(0)} \dfrac{\partial u_i^{(0)}}{\partial x_j} - \mu \Delta u_i^{(0)} = 0, i = 1,2,\cdots,n \tag{3}$$

with the irrotational condition $\dfrac{\partial u_j^{(0)}}{\partial x_i} = \dfrac{\partial u_i^{(0)}}{\partial x_j}$, $i, j = 1,2,\cdots,n$, $i \neq j$, constant $A$ and the function $\varphi = \varphi(x_1, x_2, \cdots, x_n, t)$ are to be determined later.

Substituting (2) into the left hand side of system (1), noticing that

$$\dfrac{\partial u_i}{\partial t} = A\dfrac{\partial}{\partial x_i}\left(\dfrac{\varphi_t}{\varphi}\right) + \dfrac{\partial u_i^{(0)}}{\partial t},$$

$$u_j \dfrac{\partial u_i}{\partial x_j} = \dfrac{1}{2} A^2 \dfrac{\partial}{\partial x_i}\left(\dfrac{\varphi_{x_j}}{\varphi}\right)^2 + A\dfrac{\partial}{\partial x_i}\left(u_j^{(0)} \dfrac{\varphi_{x_j}}{\varphi}\right) + u_j^{(0)} \dfrac{\partial u_i^{(0)}}{\partial x_j},$$

$$\Delta u_i = \dfrac{\partial}{\partial x_i}\left(\dfrac{\Delta\varphi}{\varphi} - \sum_{j=1}^{n}\left(\dfrac{\varphi_{x_j}}{\varphi}\right)^2\right) + \Delta u_i^{(0)},$$

and using (3), we have

$$\frac{\partial u_i}{\partial t}+\sum_{j=1}^{n}u_j\frac{\partial u_i}{\partial x_j}-\mu\Delta u_i = A\frac{\partial}{\partial x_i}\left[\frac{\varphi_t+\sum_{j=1}^{n}u_j^{(0)}\frac{\partial\varphi}{\partial x_j}-\mu\Delta\varphi}{\varphi}+\left(\frac{1}{2}A+\mu\right)\left(\sum_{j=1}^{n}\left(\frac{\varphi_{x_j}}{\varphi}\right)^2\right)\right],$$

$i = 1, 2, \cdots, n$. (4)

In order to find $A$, setting the coefficient of $\sum_{j=1}^{n}\left(\frac{\varphi_{x_j}}{\varphi}\right)^2$ to zero, yields

$\frac{1}{2}A+\mu=0$, which implies that $A=-2\mu$. (5)

Using (5) the expression (2) becomes

$$u_i = -2\mu\frac{\varphi_{x_i}}{\varphi}+u_i^{(0)}, i=1,2,\cdots,n \tag{6}$$

and the expression(4) becomes

$$\frac{\partial u_i}{\partial t}+\sum_{j=1}^{n}u_j\frac{\partial u_i}{\partial x_j}-\mu\Delta u_i = -2\mu\frac{\partial}{\partial x_i}\left[\frac{\varphi_t+\sum_{j=1}^{n}u_j^{(0)}\frac{\partial\varphi}{\partial x_j}-\mu\Delta\varphi}{\varphi}\right]=0,\; i=1,2,...,n \tag{7}$$

provided the function $\varphi = \varphi(x_1, x_2, \cdots, x_n, t)$ satifies the equation

$$\varphi_t + \sum_{j=1}^{n}u_j^{(0)}\frac{\partial\varphi}{\partial x_j}-\mu\Delta\varphi = 0. \tag{8}$$

Based on (6), (7) and (8), we come to the important conclusion that if $u_i^{(0)}=u_i^{(0)}(x_1,x_2,\cdots,x_n,t), i=1,2,\cdots,n$, be a particular solution of the n-dimensional Burgers system (1) and the function $\varphi = \varphi(x_1, x_2, \cdots, x_n, t)$ be a nonzero solution of Eq.(8), then the nonlinear transformation (6) satisties exactly the n-dimensional Burgers system (1). The transformation (6) with Eq.(8) is called a BT of the n-dimensional Burgers system (1).

In order to find exact solution of the n-dimensional Burgers system (1), we consider two particular cases of the n-dimensional BT:

**The first case is that**

When $u_i^{(0)}=0, i=1,2,\cdots,n$, then (6), (7) and (8) become

$$u_i = -2\mu\frac{\varphi_{x_i}}{\varphi}, i=1,2,\cdots,n, \tag{6$'$}$$

$$\frac{\partial u_i}{\partial t}+\sum_{j=1}^{n}u_j\frac{\partial u_i}{\partial x_j}-\mu\Delta u_i = -2\mu\frac{\partial}{\partial x_i}\left[\frac{\varphi_t-\mu\Delta\varphi}{\varphi}\right]=0, i=1,2,\cdots,n, \tag{7$'$}$$

$$\varphi_t - \mu\Delta\varphi = 0, \tag{8}'$$

respectively.

Based on (6)′, (7)′ and (8)′, we can say that if $\varphi = \varphi(x_1, x_2, \cdots, x_n, t)$ is a solution of the n-dimensional linear heat equation (8)′, then the n-dimensional nonlinear transformation (6)′ satifies exactly the n-dimensional Burgers system (1). Thus (6)′ with (8)′ have made up the n-dimensional Cole-Hopf transformation for the n-dimensional Burgers system (1), which coincides with that obtained in paper [5].

**The second case is that**

When $\varphi = u_i^{(0)}, i = 1, 2, \cdots, n$ then (6), (7) and (8) become

$$u_i = -2\mu\frac{\left(u_i^{(0)}\right)_{x_i}}{u_i^{(0)}} + u_i^{(0)}, i = 1, 2, \cdots, n. \tag{6''}$$

$$\frac{\partial u_i}{\partial t} + \sum_{j=1}^{n} u_j \frac{\partial u_i}{\partial x_j} - \mu\Delta u_i = -2\mu\frac{\partial}{\partial x_i}\left[\frac{\frac{\partial u_i^{(0)}}{\partial t} + \sum_{j=1}^{n} u_j^{(0)} \frac{\partial u_i^{(0)}}{\partial x_j} - \mu\Delta u_i^{(0)}}{\varphi}\right] = 0, i = 1, 2, \cdots, n. \tag{7''}$$

$$\frac{\partial u_i^{(0)}}{\partial t} + \sum_{j=1}^{n} u_j^{(0)} \frac{\partial u_i^{(0)}}{\partial x_j} - \mu\Delta u_i^{(0)} = 0, i = 1, 2, \cdots, n. \tag{8''}$$

respectively.

Based on (6)″, (7)″ and (8)″, we can say that if $u_i^{(0)} = u_i^{(0)}(x_1, x_2, \cdots, x_n, t)$, $i = 1, 2, \cdots, n$, is a given particular solution of n-dimensional Burgers system (1), then the n-dimensional nonlinear transformation (6)″ is another solution of it. Thus (6)″ with (8)″ have made up the n-dimensional Auto-BT for the n-dimensional Burgers system (1).

In the next section, we will use the Auto-BT to get more solutions of the n-dimensional Burgers system (1), when a particular solution of the system is given.

## 3. Application of Auto-BT

It is well-known that the n-dimensional linear heat equation (8)′ has a solution

$$\varphi(x_1, x_2, \cdots, x_n, t) = \frac{1}{(2\sqrt{\pi\mu t})^n}\exp\left[-\frac{\sum_{j=1}^{n} x_j^2}{4\mu t}\right], \tag{9}$$

which is denoted by

$$E(x_1, x_2, \cdots, x_n, t) = \frac{1}{(2\sqrt{\pi\mu t})^n} \exp\left[-\frac{\sum_{j=1}^{n} x_j^2}{4\mu t}\right].$$

Substituting (9) into the n-dimensional Cole-Hopf transformation (6)′, we have an exact solution of n-dimensional Burgers system (1) as follows

$$u_i^{(0)} = -2\mu \frac{E_{x_i}}{E} = \frac{x_i}{t}, i = 1, 2, \cdots, n, \tag{10}$$

where

$$E_{x_i} = \frac{1}{(2\sqrt{\pi\mu t})^n} \left(-\frac{x_i}{2\mu t}\right) \exp\left[-\frac{\sum_{j=1}^{n} x_j^2}{4\mu t}\right].$$

Substituting (10) which is called a seed solution into the n-dimensional auto-BT (6)″, we have another solution of n-dimensional Burgers system (1) as follows

$$u_i^{(1)} = -2\mu \frac{\left(u_i^{(0)}\right)_{x_i}}{u_i^{(0)}} + u_i^{(0)} = \frac{x_i^2 - 2\mu t}{x_i t}, i = 1, 2, \cdots, n. \tag{11}$$

Substituting (11) instead of $u_i^{(0)}$ in (6)″, into (6)″ we have another new solution of n-dimensional Burgers system (1), as follows

$$u_i^{(2)} = -2\mu \frac{\left(u_i^{(1)}\right)_{x_i}}{u_i^{(1)}} + u_i^{(1)} = \frac{x_i^3 - 6\mu x_i t}{t(x_i^2 - 2\mu t)}, \ i = 1, 2, \cdots, n. \tag{12}$$

Substituting (12) instead of $u_i^{(0)}$ in (6)″, into (6)″ we have another new solution of n-dimensional Burgers system (1), as follows

$$u_i^{(3)} = -2\mu \frac{\left(u_i^{(2)}\right)_{x_i}}{u_i^{(2)}} + u_i^{(2)} = \frac{x_i^4 - 12\mu x_i^2 t + 12\mu^2 t^2}{x_i^3 t - 6\mu x_i t^2}, \ i = 1, 2, \cdots, n. \tag{13}$$

……

and so on. The solutions $u_i^{(k)} (i = 1, 2, \ldots, n, k = 0, 1, 2, \ldots)$ are all rational functions.

In the above instance, it seems that the component $u_i$ of solution for n-dimensional Burgers system (1) depends upon the corresponding variable $x_i$ and $t$ only. For simplicity, the solutions (11), (12) and (13) are considered as the functions of variables $x, t$ and $u$ and are shown in Fig.(1).

However, we have other examples that the component $u_i$ ($i = 1, 2, \cdots, n$) of solution for n-dimensional Burgers system depends upon all variables $x_1, x_2, \cdots, x_n, t$. For example, the n-dimensional heat equation (8)' has a solution

$$\varphi = 1 + e^{\omega t}\cos\eta, \quad \omega = -\mu\left(\sum_{j=1}^{n}\alpha_j^2\right), \quad \eta = \sum_{j=1}^{n}\alpha_j x_j, \quad \alpha_j \text{-arbitrary costant.}$$

(14)

Substituting (14) into n-dimensional Cole-Hopf transformation (6)', we have a solution of n-dimensional Burgers system (1)

$$u_i^{(0)} = 2\mu\alpha_i \frac{e^{\omega t}\sin\eta}{1+e^{\omega t}\cos\eta}, \omega = -\mu\left(\sum_{j=1}^{n}\alpha_j^2\right), \eta = \sum_{j=1}^{n}\alpha_j x_j, i=1,2,\cdots,n. \tag{15}$$

Substituting (15) as a seed solution into n-dimensional Auto-BT (6)'', yields another solution of n-dimensional Burgers system (1)

$$u_i^{(1)} = 2\mu\alpha_i \frac{e^{\omega t}\sin^2\eta - \cos\eta - e^{\omega t}}{(1+e^{\omega t}\cos\eta)\sin\eta}, i=1,2,\cdots,n, \tag{16}$$

$$\omega = -\mu\left(\sum_{j=1}^{n}\alpha_j^2\right), \eta = \sum_{j=1}^{n}\alpha_j x_j, \alpha_j\text{-arbitrary costants.}$$

The component $u_i^{(1)}$ in (16) depends upon all variables $x_1, x_2, ..., x_n$ and $t$; and it is a periodic function in variable $\eta$, which is a linear combination of space variables ($x_1, x_2, ..., x_n$). For simplicity, the solutions (15) and (16) are considered as the functions of variables $\eta, t$ and $u$ and are shown in Fig.(2).

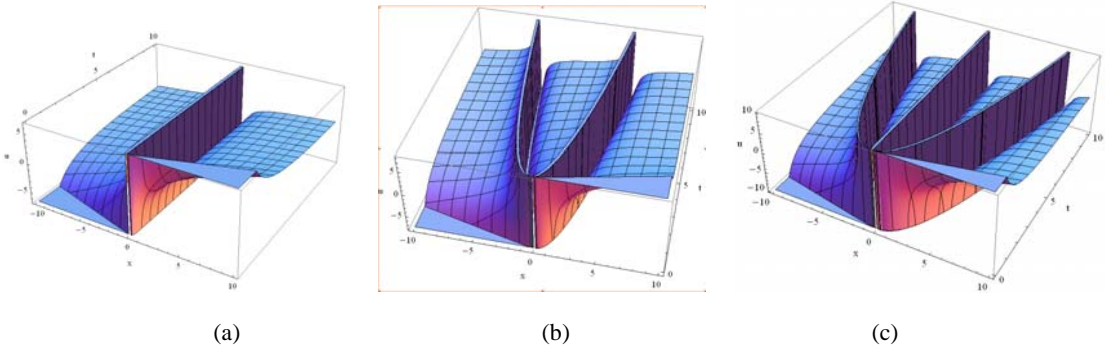

(a)　　　　　　　　(b)　　　　　　　　(c)

Fig. 1. Plots of the solutions (11), (12) and (13) with $\mu = 1$

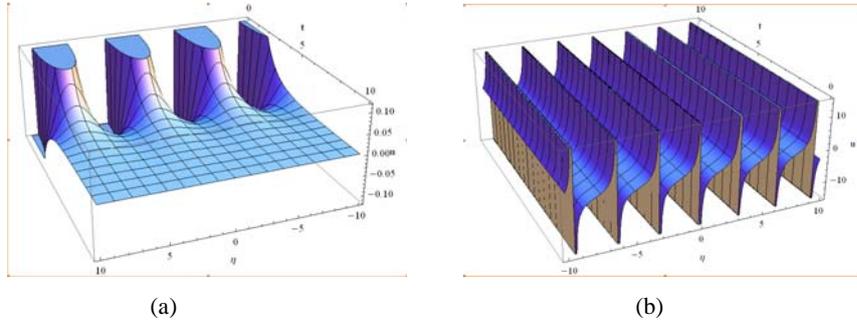

(a)　　　　　　　　(b)

Fig. 2. Plots of the solutions (15) and (16) with $\mu = 1, \omega = -1, \alpha = 1$

Similarly, we have a solution of n-dimensional Burgers system

$$u_i^{(0)} = -2\mu\alpha_i \frac{e^{\omega_1 t}\sinh\eta_1}{1+e^{\omega_1 t}\cosh\eta_1}, \quad i=1,2,...,n. \tag{17}$$

where $\omega_1 = \mu\left(\sum_{j=1}^{n}\alpha_j^2\right), \eta_1 = \sum_{j=1}^{n}\alpha_j x_j$.

Substituting (17) as a seed solution into n-dimensional Auto-BT (6)″ yields a new solution of n-dimension Burgers system (1)

$$u_i^{(1)} = -2\mu\alpha_i \frac{e^{\omega_1 t}\sinh^2\eta_1 + \cosh\eta_1 + e^{\omega_1 t}}{(1+e^{\omega_1 t}\cosh\eta_1)\sinh\eta_1}, \quad i=1,2,\cdots,n, \tag{18}$$

$\omega_1 = \mu\left(\sum_{j=1}^{n}\alpha_j^2\right), \eta_1 = \sum_{j=1}^{n}\alpha_j x_j$, $\alpha_j$-arbitrary costants.

The component $u_i^{(1)}$ in (18) depends upon all variables $x_1, x_2, ..., x_n$ and $t$ as well, for simplicity, the solutions (15) and (16) are considered as the functions of variables $\eta, t$ and $u$ and are shown in Fig.(3).

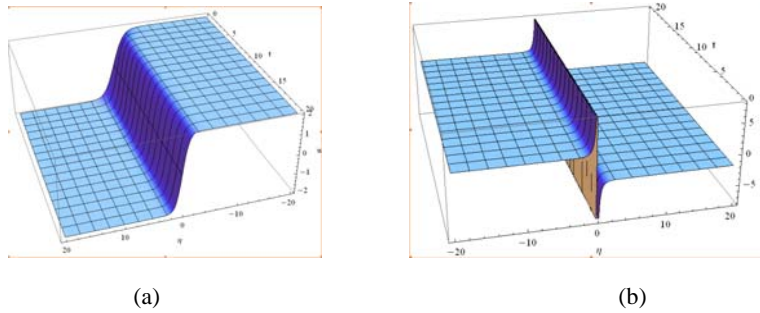

(a)          (b)

Fig. 3. Plots of the solutions (17) and (18) with $\mu=1, \omega=1, \alpha=1$.

## 4. Conclusion

Using the SHB we have derived out the n-dimensional BT for the n-dimensional Burgers system. The BT involves a given particular solution $u_i^{(0)}(i=1,2,...n)$ of the n-dimensional Burgers system as well as the solution $\phi = \phi(x_1, x_2, ..., x_n, t)$ of Eq.(8). If $u_i^{(0)}(i=1,2,...n)$ then Eq.(8) becomes classical n-dimensional heat equation, the BT becomes the well known Cole-Hopf transformation; if $u_i^{(0)} \neq 0$, and taking $\phi = u_i^{(0)}$, then Eq.(8) becomes right the n-dimension Burgers system for $u_i^{(0)}$. Thus the BT becomes the Auto-BT of the n-dimension Burgers system. Using Cole-Hopf transformation, it is easy to get a particular solution of the n-dimension Burgers system. Taking the particular solution as a seed solution of n-dimension Burgers system and using Auto-BT repeatedly, more solutions of the n-dimension Burgers system can be obtained successively.

**Acknowledgments**

The authors express their sincere thanks to the referee for valuable suggestions. This work were supported in part by the Basic Science and the Front Technology Research Foundation of Henan Province of China (Grant no. 092300410179) and the Scientific Research Innovation Ability Cultivation Foundation of Henan University of Science and Technology (Grant no.011CX011).